\def\doublespace{\baselineskip=20pt plus 3pt\message{double space}}
  \makeatletter
\newcommand{\sfrac}[2]{{\textstyle{#1\over#2}}}
\def\artsectnumbering{%
  \@addtoreset{equation}{section}
  \def\theequation{\thesection.\arabic{equation}}}
  \makeatother
\documentstyle{article}
\title{A Consistency Condition for the Double Series Approximation Method.}
\author{M.S. Piper\thanks{e-mail m.s.piper@qmw.ac.uk}
                   \\ School of Mathematical Sciences, \\
                   Queen Mary and Westfield College,\\
                   Mile End Road,\\ 
                   London.\\
                   E1 4NS}
\begin{document}
\maketitle
\begin{abstract}
The double series approximation method of Bonnor is a means for examining the 
gravitational radiation from an axisymmetric isolated source that undergoes a 
finite period of oscillation. It involves an expansion of the metric as a 
double Taylor series. Here we examine the integration procedure that is used 
to form an algorithmic solution to the field equations and point out the 
possibility of the expansion method breaking down and predicting a singularity 
along the axis of symmetry. We derive a condition on the solutions obtained 
by the double series method that must be satisfied to avoid this singularity. 
We then consider a source with only a quadrupole moment and verify that to 
fourth order in each of the expansion parameters, this 
condition is satisfied. This is a reassuring test of the consistency of the 
expansion procedure. We do, however, find that the imposition of this 
condition makes a physical interpretation of any but the lowest order 
solutions very difficult. The most obvious decomposition of the solution into 
a series of independent physical effects is shown not to be valid.
\end{abstract}
\doublespace
\artsectnumbering
\section{Introduction.}
In 1959 Bonnor introduced the doubles series method as a means for looking at
the gravitational radiation from an isolated source that oscillates for a
finite period \cite{B59}. This involved expanding the metric as a double power
series in two parameters and solving the vacuum Einstein field equations by 
successive approximations. This first approach dealt with isolated sources 
emitting spherical gravitational waves 
and showed that the source lost mass at a rate equal to that of energy 
radiated. Following the publication of the Bondi metric in 1960 \cite{Bondi}, 
a metric well suited to this problem, Bonnor and Rotenberg refined and 
generalised the double series method to the axisymmetric case \cite{BR66}. 
They were then able to reproduce the earlier result concerning mass loss and to
show that the momentum generated in the source is equal and opposite to that 
removed by the gravitational waves. They were also able to assert the existence
of wave-tails in Bondi coordinates. These represent the backscattering of the 
gravitational radiation by the spacetime curvature induced by the source. 
The tails were interpreted as incoming radiation in a further paper on the 
double series method by Hunter and Rotenberg \cite{HR69}. Elsewhere,
\cite{32soln}, I considered the interaction of these incoming wave-tails with
the Schwarzschild source. I found that the wave-tails have no permanent
effects on the source - all changes in the metric die off at least like 
$u^{-2}$ ($u$ being retarded time). I also showed that the incoming radiation 
does not alter the mass of the source itself.

The double series method itself has influenced the formalism of Blanchet, 
Damour and Iyer \cite{BDI}, in which the spacetime is divided into a near-zone 
region, a far-zone region and a region of overlap. In the far-zone a 
post-Minkowskian expansion is undertaken that is a generalisation of the double
series method to more physical, non-axisymmetric sources. A review of the BDI 
formalism has recently been given by Blanchet \cite{Blanchet}.

In this paper I derive a consistency condition for the double series method. 
If this condition is not satisfied at a given level of approximation, a 
singularity appears in the solution along the axis of symmetry. While it might 
be possible to interpret this singularity as an infinite pipe or strut, this 
would go against our original criterion imposed here, namely that the system 
be isolated. The occurrence of this singularity would therefore represent a 
breakdown of the double series method applied to isolated sources. The 
importance of the condition 
lies in the fact that at a given order of approximation its satisfaction 
depends not on the solution of the field equations to that order, but on the 
previously obtained lower order solutions. The condition is more a 
test of the double series method itself than of any particular solution 
obtained by the method.

We then extend the work on the double series method using a power series 
package developed for SHEEP to carry out the otherwise prohibitive algebraic 
calculations \cite{ps}. We verify that the consistency condition is satisfied
at the third order and obtain the corresponding solution. We are then able to
show that to fourth order the consistency condition is also satisfied. 
However, when considering specifically the wave-tail - wave-tail interaction 
we find that the condition fails. This component of the solution generates 
terms in the metric that are singular along the axis of symmetry. We are able 
to show that there is an exact cancellation that removes these singular terms 
in the full solution. This suggests that the double series method is not 
decomposable 
in the sense that it cannot be easily separated into a series of individual 
physical effects but that there is a complicated mixing of these effects 
occurring. This in turn leads to difficulties in the physical interpretation 
of the solutions beyond the lowest orders.

The order of this paper is as follows. In section 2 we describe the source
and the expansion of the metric coefficients that is characteristic of the
double series method. In section 3 we outline the algorithm that is used to
solve the linearised field equations and in section 4 we describe how this
algorithm leads to the possible occurrence of singularities along the axis
of symmetry. We then derive the condition for the avoidance of these 
singularities. Section 5 reviews previous work on the double series method and
 verifies that the solutions obtained satisfy this consistency condition. In 
sections 6 and 7 we extend the solution, for a source with only a quadrupole
moment, to the third and fourth orders and 
verify that these solutions are consistent with our assumption of an isolated 
source to fourth order.

\section{The Double Series Method.}
We consider an isolated axisymmetric source, such as two particles oscillating
along an axis. The system is at rest for $u<u_{1}$, vibrates smoothly for the
period $u_{1}<u<u_{2}$ and is again at rest for $u>u_{2}$, though not 
necessarily in the same state as before $u_{1}$. We are interested in
the behaviour of the system at large radial distances and so do not specify the
source in detail. We do not describe the mechanism that drives the 
oscillations since this should not have an effect on the asymptotic behaviour
of the solution.

It is assumed that there are two characteristic measurements
of the system, $m$, the source mass, and $a$, a characteristic length
(the separation of the two particles in the example mentioned above).
The fundamental assumption of the double series method is that there exist
solutions of the vacuum field equations that admit Taylor expansions in $m$
and $a$ about $m=0$, $a=0$. We do not offer a proof that this is the case, but
the linear approximation suggests that this is a valid assumption to make.

We assume that the system can be modelled by Bondi's radiative metric 
\cite{Bondi} which can be written in the form:
\begin{equation}
ds^{2} = -r^{2}(Bd\theta^{2}+C\sin^{2}\theta d\phi^{2})+Ddu^{2}+2Fdrdu
+2rGd\theta du
\end{equation}
with $C=B^{-1}$ and where the metric coefficients are functions of $u,r,\theta$
. Here we are using Bondi coordinates labelled 
$(x_{1},x_{2},x_{3},x_{4})=(r,\theta,\phi,u)$.

We make two further assumptions in order to start the approximation procedure.
The first is that if we remove the mass $(m=0)$ our metric should reduce to
flat spacetime in Bondi's coordinates:
\begin{equation}
ds^{2} = -r^{2}(d\theta^{2}+\sin^{2}\theta d\phi^{2})+du^{2}+2drdu.
\label{eq:flat}
\end{equation}
The second assumption is that if we set $a=0$ then the spacetime should reduce
to Schwarzschild space in Bondi coordinates:
\begin{equation}
ds^{2} = -r^{2}(d\theta^{2}+\sin^{2}\theta d\phi^{2})+(1-\frac{2m}{r})du^{2}
+2drdu.
\label{eq:schwar}
\end{equation}

Our fundamental assumption leads to the following representation:
\begin{eqnarray}
B & = & 1+\sum_{p=1}^{\infty} \sum_{s=0}^{\infty}\stackrel{(ps)}{B}m^{p}a^{s}
\nonumber \\
C & = & 1+\sum_{p=1}^{\infty} \sum_{s=0}^{\infty}\stackrel{(ps)}{C}m^{p}a^{s}
\nonumber \\
D & = & 1+\sum_{p=1}^{\infty} \sum_{s=0}^{\infty}\stackrel{(ps)}{D}m^{p}a^{s}
\nonumber \\
F & = & 1+\sum_{p=1}^{\infty} \sum_{s=0}^{\infty}\stackrel{(ps)}{F}m^{p}a^{s}
\nonumber \\
G & = & \sum_{p=1}^{\infty} \sum_{s=0}^{\infty}\stackrel{(ps)}{G}m^{p}a^{s}.
\label{eq:expansion}
\end{eqnarray}
To zeroth order in $m$ this corresponds to the metric (\ref{eq:flat}). The
choice $\stackrel{(10)}{D}=-2r^{-1}$, with all other $\stackrel{(s0)}{B}$,
$\stackrel{(s0)}{C}$, $\stackrel{(s0)}{D}$, $\stackrel{(s0)}{F}$,
$\stackrel{(s0)}{G}$, $(s\geq0)$ vanishing,
ensures that the metric reduces to (\ref{eq:schwar}) to zeroth order in $a$. 
Here again the coefficients in the expansion (\ref{eq:expansion}) are functions
of $u,r,\theta$. We now introduce the notation $\stackrel{(ps)}{g_{ab}}$ to
represent the terms in the metric proportional to $m^{p}a^{s}$. By the $(ps)$
case we shall mean the solution of the field equations to order $m^{p}a^{s}$. 
Where there is ambiguity as to which products give rise to $m^{p}a^{s}$ terms
(for example in the $(24)$ case we have
\begin{equation}
ma^{2} \times ma^{2} , m \times ma^{4} , ma \times ma^{3}
\label{eq:24prod}
\end{equation}
terms), we shall refer to these products explicitly, e.g. the $(12)\times (12)$
case, the $(10)\times (14)$ case or the $(11)\times (13)$ case in the example 
above.
 
Having made the expansion (\ref{eq:expansion}) above, the vacuum field 
equations are of the form:
\begin{equation}
\Phi(\stackrel{(ps)}{g_{ab}})=\Psi(\stackrel{(qr)}{g_{ab}})
\label{eq:EFE}
\end{equation}
where the left hand side is linear in $\stackrel{(ps)}{g_{ab}}$ (the unknowns)
and the right hand side is non-linear in $\stackrel{(qr)}{g_{ab}}$, $q<p$, 
$r<s$ which are known from previous steps in the approximation. The explicit 
form of these equations is given in the appendix.

We have now arrived at a system of linear partial differential equations.
In the next section we describe the algorithm that is used to obtain
solutions to this system.

\section{The Algorithm for Solving the Field Equations.}

It can be shown that the field equations (\ref{eq:EFE}), three of which are
identically satisfied due to our imposition of axial symmetry, admit an 
algorithmic solution. The derivation of this
algorithm is given in the appendix. Here we give the bare details that are
required for the derivation of the consistency condition. In order to ease
the notation we consider the general $m^{p}a^{s}$ case and omit the 
superscripts $(ps)$. The algorithm consists of three steps:

{\bf Step 1.}

Obtain $F$, the coefficient of $g_{12}$ of order $m^{p}a^{s}$, as defined 
by (\ref{eq:expansion}) by the relationship:

\begin{equation}
F=-\sfrac{1}{4}\int r H dr+\eta(\theta,u)
\label{eq:step1}
\end{equation}
where $\eta$ is a function of integration and 
$H$ is one of the non-linear terms, i.e. one component of the 
right-hand side of equations (\ref{eq:EFE}).

{\bf Step 2.}

Solve for $D$ the `pseudo-wave' equation
\begin{eqnarray}
\Box^{'}D&\equiv&D_{11}-2D_{14}+2r^{-1}(D_{1}+D_{4})+r^{-2}(D_{22}
+D_{2}\cot\theta)
\nonumber \\
&=&-K+2(F_{14}+2r^{-1}F_{4})+\nonumber \\
& &2r^{-2}\left[\,\int\{r^{2}(N-2F_{14})+(F_{22}+F_{2}
\cot\theta)\}dr+\chi(u,\theta)\,\right]_{4}
\label{eq:pseudo}
\end{eqnarray}
where $K$, $N$ are non-linear terms,
$\chi$ is a function of integration and $F$
is defined by (\ref{eq:step1}). The pseudo-wave operator denoted $\Box^{'}$
is similar to the usual d'Alembertian wave operator of flat space in the Bondi
coordinates used here, the only difference coming in the sign of the
$2D_{4}r^{-1}$ term.

{\bf Step 3.}

Obtain the remaining metric coefficients from the expressions
\begin{eqnarray}
G&=&
r^{-1}{\rm cosec}\,\theta\int\sin\theta\left[\,\int r^{2}
(N-2F_{14})dr+r^{2}D_{1}+\chi\,\right]d\theta\nonumber \\ & &
+r^{-1}\int F_{2}dr+\nu(r,u)\,{\rm cosec}\,
\theta
\label{eq:Geqn}
\end{eqnarray}
\begin{eqnarray}
B&=&{\rm cosec}^{2}\,\theta\int\sin^{2}\theta\left[-\int\{rL
+2r^{-1}(F_{2}-G)\}dr
+F_{2}-G-rG_{1}\,\right]d\theta\nonumber \\& &+\tau(r,u)\,
{\rm cosec}^{2}\,\theta+
\mu(\theta,u)
\label{eq:Beqn}
\end{eqnarray}
where $\nu$, $\tau$, $\mu$
are further functions of integration and $L$ is another
non-linear term.

The five functions of integration that arise in the double series method,
namely $\eta$,$\chi$,$\nu$,$\tau$ and $\mu$, are chosen to satisfy two
criteria. Firstly they should ensure that the solution satisfies the full set
of field equations (\ref{eq:EFE}) (the derivation of the above algorithm
depends on only 4 of the 7 non-trivial field equations). Secondly they are used
to impose certain boundary conditions on the system, namely that the metric be 
Minkowskian at spatial infinity, and that the metric remain non-singular on
the axis of symmetry except at the origin. A sufficient condition for the
latter is that $B\,{\rm cosec}^{2}\,\theta, C\,{\rm cosec}^{2}\,\theta, D,$ 
$F, G\,{\rm cosec}\,\theta$ are of class 
$C^{2}$ near $\sin\theta=0$. In the next section we show that unless a
particular condition is satisfied this criterion cannot be achieved. This 
imposes a consistency condition on the double series method.

\section{A Consistency Condition.}
We now examine the possibility of a singularity arising along the axis of 
symmetry during the integration of the field equations contained in the
algorithm of the previous section. This singularity cannot be reconciled with
our assumption of an isolated source and its existence would imply a lack of 
consistency in this application of the double
series method. In later sections we will show that this condition is indeed
satisfied to order $m^{4}a^{4}$. Here we derive the condition itself. As in
the previous section we consider the general $m^{p}a^{s}$ case and drop the 
superscripts $(ps)$ to ease the notation.

We suppose that we have obtained a solution $F$ of 
(\ref{eq:step1}) and a solution $D$ of the psuedo-wave 
equation (\ref{eq:pseudo}). Since the wave operator $\Box^{'}$ defined by 
(\ref{eq:pseudo}) preserves the Legendre polynomials in $\cos\theta$, it is 
natural to make the following expansion of the theta dependence of the 
solution:
\begin{eqnarray}
D & = & \sum_{i=0}^{\infty} \,_{i} D(u,r) P_{i}(\cos\theta),
\ \ F = \sum_{i=0}^{\infty} \,_{i} F(u,r) P_{i}(\cos\theta)
\nonumber \\
N & = & \sum_{i=0}^{\infty} \,_{i} N(u,r) P_{i}(\cos\theta),
\ \ \chi = \sum_{i=0}^{\infty} \,_{i} \chi (u) P_{i}(\cos\theta).
\label{eq:thetaexp}
\end{eqnarray}
where $P_{i}$ is the $i$th Legendre polynomial.

We now proceed to step 3 of the algorithm and obtain an expression for $G$
from (\ref{eq:Geqn}):
\begin{eqnarray}
G&=&\sum_{i=0}^{\infty} r^{-1}{\rm cosec}\,\theta\int\sin\theta 
P_{i}(\cos\theta)\left[\,\int r^{2}(\,_{i}N-2\,_{i}F_{14})dr+r^{2}\,_{i}D_{1}
+\,_{i}\chi\,\right]d\theta\nonumber \\ & &+\sum_{i=0}^{\infty} 
r^{-1}P_{i}(\cos\theta)^{'}\int \,_{i}Fdr+\nu(u,r)\,{\rm cosec}\,\theta,
\label{eq:gexp}
\end{eqnarray}
where $\,^{'}$ here denotes differentiation with respect to theta.
Let us now take the first term $(i=0)$ of the series in the above expansion.
This is
\begin{equation}
-r^{-1}\cot\theta\left[\,\int r^{2}(\,_{0}N-2\,_{0}F_{14})dr+r^{2}\,_{0}D_{1}
+\,_{0}\chi\,\right]
\label{eq:p0term}
\end{equation}
which is singular at $\sin\theta=0$, i.e. along the axis of symmetry. Since
\begin{equation}
\int P_{i}(\cos\theta)\sin\theta d\theta
\end{equation}
contains a factor $\sin\theta$ for $i\geq1$, the expression (\ref{eq:p0term})
is the only singular term in the series contained in (\ref{eq:gexp}). Moreover,
this singularity cannot be removed by a specific choice of the function of
integration $\nu$ in (\ref{eq:gexp}). The choice 
\begin{equation}
\nu = \pm r^{-1}\left[\,\int r^{2}(\,_{0}N-2\,_{0}F_{14})dr+r^{2}\,_{0}D_{1}
+\,_{0}\chi\,\right]
\label{eq:mu}
\end{equation}
removes the singularity along either the semi-axis $\theta = 0$, or the 
semi-axis $\theta = \pi$ but not both. Thus we find that the double series
method predicts a singularity along the axis of symmetry unless the term
(\ref{eq:p0term}) vanishes, i.e. unless
\begin{equation}
\,_{0}D_{1} = -r^{-2}\int r^{2}(\,_{0}N-2\,_{0}F_{14})dr-r^{-2}\,_{0}\chi.
\label{eq:cond1}
\end{equation}
We now take this expression for $\,_{0}D_{1}$ and substitute it into the 
pseudo-wave equation (\ref{eq:pseudo}) which leads us to an expression for
$\,_{0}D_{4}$:
\begin{equation}
\,_{0}D_{4} = r\left[\,\sfrac{1}{2}\,_{0}N-\sfrac{1}{2}\,_{0}K\,\right]
+2\,_{0}F_{4}.
\label{eq:cond2}
\end{equation}
Now using the fact that partial derivatives commute we can obtain two 
expressions for $\,_{0}D_{14}$ from (\ref{eq:cond1}) and (\ref{eq:cond2}) 
which must be equivalent. This leads us to the following condition that must 
be satisfied if a singularity along the axis is to be avoided:
\begin{eqnarray}
X & := & r^{2}\left[\,\sfrac{1}{2}\,_{0}N_{11}-\sfrac{1}{2}\,_{0}K_{11}
+\sfrac{1}{2}\,_{0}H_{44}-\sfrac{1}{2}\,_{0}H_{14}\,\right] +\nonumber \\
& & r\left[\,2\,_{0}N_{1}-2\,_{0}K_{1}+\,_{0}N_{4}
-\sfrac{3}{2}\,_{0}H_{4}\,\right]+\left[\,_{0}N-\,_{0}K \,\right]\nonumber \\
&\equiv& 0.
\label{eq:cond}
\end{eqnarray}
This is a necessary condition for the avoidance of singularities at the $(ps)$
level. It is also a sufficient condition for the avoidance of singularities in 
$\stackrel{(ps)}{G}$ in the sense that if (\ref{eq:cond})
is satisfied then a particular choice of the functions of integration $\chi$,
 $\eta$ can always be made to ensure that there is no singularity along the
axis in $\stackrel{(ps)}{G}$. There is the additional possibility of the 
integration procedure leading to a singularity along the axis in 
$\stackrel{(ps)}{B}$. The avoidance of this singularity leads to a condition 
on the $P_{1}(\cos\theta)$ 
coefficients of the non-linear terms and on $\,_{0}D_{1}$. However, this 
condition turns out not to be as simple as (\ref{eq:cond}), nor is it relevant 
to the rest of the work in this paper as we shall see in the next section. 

It can now be seen that the condition (\ref{eq:cond}) depends only on the 
non-linear terms $\,_{0}N$, $\,_{0}H$, $\,_{0}K$, i.e. only on the metric
coefficients $\stackrel{(qr)}{{g}_{ab}}$, $q<p$, $r<s$, that have been found by
previous approximations. The occurrence of a singularity along the axis would,
for the type of sources we have in mind here, appear unphysical. Any 
interpretation of this singularity as a strut or pipe extending along an 
infinite semi-axis would contradict our assumption the the source be isolated. 
Therefore in solving the $(qr)$, $q<p$, $r<s$ approximations we require a
relationship between these solutions to be satisfied at the $(ps)$ order. This
relationship does not seem to be manifestly satisfied. We therefore conclude
that the condition (\ref{eq:cond}) is a test of the double series method 
itself. If (\ref{eq:cond}) is not satisfied at any particular level of 
approximation the expansion procedure does not seem to be consistent with our
premise of having an isolated source. We now examine the work of Bonnor and 
Rotenberg 
\cite{BR66} on the double series method and show that the solutions obtained 
are consistent in the above sense.

\section{The First and Second Order Solutions.}
We now briefly review the first and second order (in $m$) solutions obtained by
the double series method, paying particular attention to the condition 
(\ref{eq:cond}) derived in the previous section. The first and second order 
solutions were found and interpreted by Bonnor and Rotenberg \cite{BR66}.

The linear approximation, $(1s)$, is a superposition of terms each involving
one of the multipole moments, $Q_{n}$, about the axis of symmetry. We use the
non-gravitational forces inherent in the system to ensure that these have a
suitable form:
\begin{equation}
Q_{n}=ma^{n}h_{n}(u)
\label{eq:moments}
\end{equation}
for $n>1$. The dipole moment can be eliminated by a suitable choice of 
frame. We choose such a frame and then are able to specify that 
$\stackrel{(11)}{g_{ab}}=0$. The functions $h_{n}$ describe the oscillation of 
the source. Here, for simplicity we assume that only $h_{2}$ is not 
identically zero so that the source has only a quadrupole moment. This ensures 
that $\stackrel{(ps)}{H}$, $\stackrel{(ps)}{K}$, $\stackrel{(ps)}{N}$, 
$\stackrel{(ps)}{D}$ all have only even order dependence on the Legendre 
polynomials in $\cos\theta$ and thay $\stackrel{(ps)}{L}$ depends only on the 
derivatives of even order Legendre polynomials in $\cos\theta$. As a result 
the condition mentioned in the 
previous section for the avoidance of singularities in $\stackrel{(ps)}{B}$ 
will always be trivially satisfied in this work.
The function $h_{2}$ is assumed to be
constant for $u<u_{1}$, $u>u_{2}$ and to vary smoothly for $u_{1}<u<u_{2}$.
In the double series method all source terms are inserted at the linear
approximation. This is a distinct advantage over the Bondi, van den Burg and
Metzner formalism,(\cite{BBM61}), in which the news function that generates 
the solution
contains non-linear terms. To ensure that we do not insert higher order source 
terms we specify that any arbitrary functions of integration at higher orders
are set to zero, except where these are required to avoid a singularity 
occurring along the axis of symmetry.

Bonnor and Rotenberg then solved the $(22)$ approximation. The non-linear 
terms that arise at the $m^{2}a^{2}$ level are due to 
\begin{equation}
\stackrel{(10)}{g_{ab}} \times \stackrel{(12)}{g_{ab}}
\end{equation}
products. The $(22)$ solution thus describes the interaction between the 
Schwarzschild source and the quadrupole wave. The solution contains integrals 
of the form:
\begin{equation}
S_{n}=\int^{r}_{\infty}w^{-n}h_{2}(u+2r-2w)dw.
\label{eq:tail}
\end{equation}
These represent wave-tails, describing the back-scattering of the quadrupole 
radiation by the source.

The $(24)$ solution was solved to $O(r^{-3})$ by Bonnor and Rotenberg and 
completed by Hunter and Rotenberg \cite{HR69}. It is here that the loss of 
mass of the source due to the gravitational radiation is first seen. The 
occurrence of
such integrals as:
\begin{equation}
Y:=\int^{u}_{-\infty}\stackrel{...}{h_{2}}^{2}du
\label{eq:mloss}
\end{equation}
which are zero for $u<u_{1}$ and non-zero positive constants for $u>u_{2}$
indicate a permanent change in the metric due to the period of oscillation. It
can be shown that they represent a source losing mass at a rate equal to that
at which energy is radiated by gravitational waves.

The $(24)$ approximation is the first at which the consistency condition
(\ref{eq:cond}) is not trivially satisfied. At the $(22)$ level, $\,_{0}H$,
$\,_{0}K$, $\,_{0}N$ all vanish. In the $(24)$ approximation, Hunter and
Rotenberg found that
\begin{eqnarray}
\,_{0}H & = & -\sfrac{2}{15}r^{-4}\ddot{h_{2}}^{2} - \sfrac{4}{5}r^{-6}
\ddot{h_{2}}h_{2} - \sfrac{6}{5}r^{-8}h_{2}^{2}\nonumber \\ 
\,_{0}K & = & 
 - \sfrac{2}{15}r^{-2}\stackrel{...}{h_{2}}^{2} + r^{-4}\left[\,-
\sfrac{4}{15}\stackrel{...}{h_{2}}\dot{h_{2}} -\sfrac{4}{15}\ddot{h_{2}}^{2}
\,\right] - \sfrac{4}{15}r^{-5}\ddot{h_{2}}\dot{h_{2}}+\nonumber \\
& & r^{-6}\left[\, - \sfrac{4}{5}\ddot{h_{2}}h_{2} + 2\dot{h_{2}}^{2}\,\right]
+ \sfrac{26}{5}r^{-7}\dot{h_{2}}h_{2} + 3r^{-8}h_{2}^{2}
\nonumber \\ 
\,_{0}N & = &
\sfrac{2}{15}r^{-3}\stackrel{...}{h_{2}}\ddot{h_{2}} + r^{-5}\left[\,
\sfrac{2}{5}\stackrel{...}{h_{2}}h_{2} - \sfrac{2}{5}\ddot{h_{2}}
\dot{h_{2}}\,\right]+ \nonumber \\
& & r^{-6}\left[\, - \sfrac{6}{5}\ddot{h_{2}}h_{2} + \sfrac{8}{5}
\dot{h_{2}}^{2}\,\right] + \sfrac{26}{5}r^{-7}
\dot{h_{2}}h_{2} + 3r^{-8}h_{2}^{2}
\label{eq:24terms}
\end{eqnarray}
It can easily be checked that these values satisfy the condition. Thus to 
second order the expansion method of the double series method is seen to allow
a consistent specification of an isolated source. It should be noted that the 
non-linear terms 
(\ref{eq:24terms}) arise only from the product
\begin{equation}
\stackrel{(12)}{g_{ab}} \times \stackrel{(12)}{g_{ab}}.
\end{equation}
At the third and fourth orders it is not this simple - the non-linear terms 
consist of several such products. If the double series method were to be 
`clean', each of these products could be considered independently. This does 
not turn out to be the case.

\section{Third Order Solutions.}
The $(32)$ solution is given in another paper \cite{32soln}. Here it is 
sufficient to note that at this order the condition (\ref{eq:cond}) is 
trivially satisfied. Here we are interested in the $(34)$ solution.

The non-linear terms of order $m^{3}a^{4}$ are formed by the following 
products:
\begin{equation}
\stackrel{(10)}{g_{ab}} \times \stackrel{(24)}{g_{ab}},\ \ 
\stackrel{(12)}{g_{ab}} \times \stackrel{(22)}{g_{ab}}, \ \ 
\stackrel{(10)}{g_{ab}} \times \stackrel{(12)}{g_{ab}}\times 
\stackrel{(12)}{g_{ab}}.
\label{eq:34prod}
\end{equation}
Let us consider each case in turn. Firstly we consider those non-linear terms
that arise through $(10)\times (24)$ products. Using the power series package
developed for SHEEP \cite{ps} we can calculate the corresponding values of
$\,_{0}H$,$\,_{0}K$ and $\,_{0}N$. We do not give these values here since
they are not of particular interest in themselves. When we test the consistency
condition we find that it is not satisfied. The value of $X$ in (\ref{eq:cond})
is non-zero. We denote this value $X_{1}$:
\begin{eqnarray}
X_{1} & = & \sfrac{2}{15}r^{-4}\stackrel{...}{h_{2}}\ddot{h_{2}} + 
r^{-6}\left[\, -\sfrac{14}{5}\stackrel{...}{h_{2}}h_{2} + 
\sfrac{66}{5}\ddot{h_{2}}\dot{h_{2}}\,\right] + 
\nonumber \\& &r^{-7}\left[\,36\ddot{h_{2}}h_{2} - 48\dot{h_{2}}^{2}\,\right] 
- \sfrac{994}{5}r^{-8}\dot{h_{2}}h_{2} - 168r^{-9}h_{2}^{2}.
\label{eq:c1}
\end{eqnarray}
Now we consider the $(12)\times (22)$ case. Again we find that the consistency
condition fails. We denote the values of $X$ that we find in this case $X_{2}$:
\begin{eqnarray}
X_{2}& = & \sfrac{2}{15}r^{-4}\stackrel{...}{h_{2}}\ddot{h_{2}} - 
\sfrac{16}{5}r^{-5}\ddot{h_{2}}^{2} + r^{-6}\left[\,\sfrac{22}{5}
\stackrel{...}{h_{2}}h_{2} + \sfrac{146}{15}\ddot{h_{2}}\dot{h_{2}}\,\right] 
\nonumber \\& &- 12r^{-7}\ddot{h_{2}}h_{2} + \sfrac{342}{5}r^{-8}
\dot{h_{2}}h_{2} + \sfrac{336}{5}r^{-9}h_{2}^{2}.
\label{eq:c2}
\end{eqnarray}
Finally we look at the non-linear terms that arise through 
$(10)\times (12)\times (12)$ products. We find a third, non-zero value for $X$
which we this time denote $X_{3}$:
\begin{eqnarray}
X_{3}& = & -\sfrac{4}{15}r^{-4}\stackrel{...}{h_{2}}\ddot{h_{2}} + 
\sfrac{16}{5}r^{-5}\ddot{h_{2}}^{2} + r^{-6}\left[\,-\sfrac{8}{5}
\stackrel{...}{h_{2}}h_{2} - \sfrac{344}{15}\ddot{h_{2}}\dot{h_{2}}\,\right] 
\nonumber \\& &+ r^{-7}\left[\,-24\ddot{h_{2}}h_{2}+48\dot{h_{2}}^{2}\,\right] 
+ \sfrac{652}{5}r^{-8}\dot{h_{2}}h_{2} + \sfrac{504}{5}r^{-9}h_{2}^{2}.
\label{eq:c3}
\end{eqnarray}

Thus we find that each individual product in (\ref{eq:34prod}) fails to 
satisfy the condition (\ref{eq:cond}) and hence leads to singularities in 
the metric. However, if we consider all the products together, i.e. the 
complete $(34)$ case, we find that the condition is satisfied. This is seen 
by the relationship:
\begin{equation}
X_{1}+X_{2}+X_{3}=0.
\end{equation}
The singular terms that appear due to the non-zero values of $X$ given by 
(\ref{eq:c1}) to (\ref{eq:c3}) exactly cancel each other. Thus the double 
series method to this order is mixed in the sense that the decomposition 
of the problem into the individual cases given by (\ref{eq:34prod}) leads to a 
solution that in each case is inconsistent with our assumption that the source 
is isolated. Only the full solution is valid. We have
obtained the solution to the $(34)$ field equations in order to proceed to the
fourth order solution but we do not give it here. The fact that it cannot be
decomposed into its constituent cases would lead to great difficulties in 
making a physical interpretation of the solution. This is certainly the case 
at the $(44)$ level which we consider next.

\section{Fourth Order Solutions.}
Here we consider the $(44)$ solution. This would appear to be of significant 
physical interest since it contains the first interaction between the 
wave-tails, arising from the product
\begin{equation}
\stackrel{(22)}{g_{ab}} \times \stackrel{(22)}{g_{ab}}.
\end{equation}
However, when we calculate the non-linear terms due to this product we find 
that the consistency condition (\ref{eq:cond}) is not satisfied. In particular
\begin{eqnarray}
X &=&\sfrac{4}{5}r^{-3}\stackrel{...}{h_{2}}S_{5} + r^{-4}\left[\,
\stackrel{...}{h_{2}}S_{4} - \sfrac{12}{5}\ddot{h_{2}}S_{5}\,\right] + r^{
-5}\left[\,\sfrac{4}{5}\stackrel{...}{h_{2}}S_{3} - 8\ddot{h_{2}}S_{4} + 
\sfrac{36}{5}\dot{h_{2}}S_{5}\,\right] + \nonumber \\
& &r^{-6}\left[\, - \sfrac{1}{5}
\stackrel{...}{h_{2}}\dot{h_{2}} + \sfrac{2}{5}\stackrel{...}{h_{2}}S_{2} + 
\sfrac{2}{15}\ddot{h_{2}}^2 - \sfrac{196}{15}\ddot{h_{2}}S_{3} + 
9\dot{h_{2}}S_{4} + \sfrac{24}{5}h_{2}S_{5}\,\right] + \nonumber \\
& &r^{-7}\left[\,6\ddot{h_{2}}\dot{h_{2}} - 12
\ddot{h_{2}}S_{2} + \sfrac{36}{5}\dot{h_{2}}S_{3} - 
\sfrac{108}{5}h_{2}S_{4}\,\right] + r^{-8}\left[\,3\ddot{h_{2}}h_{2} -
 \sfrac{9}{5}\dot{h_{2}}^2 + \right. \nonumber \\
& &\left.\sfrac{18}{5}\dot{h_{2}}S_{2} - \sfrac{312}{5}h_{2}S_{3}\,\right] + 
r^{-9}\left[\,\sfrac{168}{5}\dot{h_{2}}
h_{2} - \sfrac{336}{5}h_{2}S_{2}\,\right] + \sfrac{54}{5}r^{-10}h_{2}^2
\label{eq:44cond}
\end{eqnarray}

Thus the $(22)\times (22)$ case leads to singularities along the axis 
appearing in the metric at the $m^{4}a^{4}$ level. 

However, just as at the third order, we should also consider all possible 
products that arise in the non-linear terms:
\begin{equation}
\stackrel{(10)}{g_{ab}} \times \stackrel{(34)}{g_{ab}},\ \ 
\stackrel{(12)}{g_{ab}} \times \stackrel{(32)}{g_{ab}}, \ \ 
\stackrel{(22)}{g_{ab}} \times \stackrel{(22)}{g_{ab}}
\label{eq:44prod}
\end{equation}
as well as triple and quadruple products. We then find that the condition 
(\ref{eq:cond}) is again satisfied. The double series method allows a 
consistent
definition of an isolated source to fourth order. The singular terms that do 
appear in the wave-tail - wave-tail interaction (\ref{eq:44cond}) are exactly 
cancelled in the full $(44)$ solution. This suggests that it is not possible 
to separate the physically interesting part of the $(44)$ solution from that 
part with less apparent physical interest. This would make any physical 
interpretation of the $(44)$ solution very difficult.

\section{Discussion.}
We have derived a condition for the double series method that must be 
satisfied at each order for the solution to that order to be consistent with 
the idea of an isolated source. Failure to satisfy this condition leads to a 
singularity along the axis of symmetry at that order. We have then verified 
that this condition is satisfied to order $m^{4}a^{4}$ for a source with only 
a quadrupole moment. This certainly suggests that the particular relationship 
(\ref{eq:cond}) is a feature of the field equations and that the condition is 
automatically satisfied when we are dealing with isolated sources. However it 
seems very difficult to prove this. It would appear likely that any 
generalisation of the double series method to isolated sources without 
axisymmetry would require an analogous condition to be satisfied.

When we considered the third and fourth order solutions we found that the most 
straight-forward decomposition of the solution into a series of independent 
physical effects proves to be impossible. There is a complicated mixing of 
the individual interactions that ensures that despite the fact that the 
separate cases each seem to generate singular terms in the metric, the full 
solution remains non-singular and consistent with the definition of a source 
that remains isolated. The cancellation of these singular terms that occurs 
in the full solution at each order, is certainly surprising. It is also very 
important for the validity of the expansion procedure inherent in the double 
series method.

\section{Acknowledgements.}
I am very grateful to Professor M.A.H.MacCallum for many helpful discussions
during the course of this work. I am also particularly indebted to Professor
W.B.Bonnor for many interesting suggestions and a great deal of assistance.
This work was carried out under a grant from the Engineering and Physical
Sciences Research Council.
 
\appendix
\section{Appendix}
Here we present the vacuum equations used in the double series method and give
details of solving these equations. To order $m^{p}a^{s}$ the vacuum Einstein
equations are given by the vanishing of the Ricci tensor components which are
(dropping the labels $\stackrel{(ps)}{ }$):
\begin{eqnarray}
2R_{11}&\equiv&-4r^{-1}F_{1}-H \nonumber \\
2r^{-2}R_{22}&\equiv&B_{11}-2B_{14}+2r^{-1}(B_{1}-B_{4}+D_{1}-F_{1}-G_{12})
+\nonumber \\ & &r^{-2}(-B_{22}-3B_{2}\cot\theta+2B+2D+2F_{22}-4F-4G_{2}-
\nonumber \\& &\ \ \ \ \ \ 2G\cot\theta)-I\nonumber \\
-2R^{3}_{3}&\equiv&-B_{11}+2B_{14}+2r^{-1}(-B_{1}+B_{4}+D_{1}-F_{1}-G_{1}\cot
\theta)+\nonumber \\ & &r^{-2}(-B_{22}-3B_{2}\cot\theta+2B+2D+2F_{2}\cot\theta
-4F-\nonumber \\& &\ \ \ \ \ \ 2G_{2}-4G\cot\theta)-J \nonumber \\
2R_{44}&\equiv&-D_{11}+2F_{14}+2r^{-1}(-D_{1}-D_{4}+2F_{4}+G_{24}+G_{4}\cot
\theta)-\nonumber \\ & &r^{-2}(D_{22}+D_{2}\cot\theta)-K \nonumber \\
2r^{-1}R_{12}&\equiv&-G_{11}+r^{-1}(-B_{12}-2B_{1}\cot\theta+F_{12}-2G_{1})+
\nonumber \\ & &2r^{-2}(-F_{2}+G)-L\nonumber \\
2R_{14}&\equiv&-D_{11}+2F_{14}+r^{-1}(-2D_{1}+G_{12}+G_{1}\cot\theta)+
\nonumber \\ & &r^{-2}(-F_{22}-F_{2}\cot\theta+G_{2}+G\cot\theta)-N \nonumber 
\\
2r^{-1}R_{24}&\equiv&-G_{11}+G_{14}+r^{-1}(-B_{24}-2B_{4}\cot\theta
-D_{12}+F_{12}+F_{24}-\nonumber \\& &2G_{1}-G_{4})-P
\end{eqnarray}
where $H, I, J, K, L, N, P$ are the non-linear parts of the equations
(the right hand sides of (\ref{eq:EFE})).

The first equation is directly integrable with respect to $r$ to give:
\begin{equation}
F=-\sfrac{1}{4}\int r H dr+\eta(\theta,u)
\end{equation}
where $\eta$ is a function of integration.
We can also integrate the $R_{14}$ component with respect to $r$ to give:
\begin{equation}
r(G_{2}+G\cot\theta)=r^{2}D_{1}+\int\{r^{2}(N-2F_{14})+(F_{22}+F_{2}\cot\theta)
\}dr+\chi(\theta,u)
\label{eq:geqnapp}
\end{equation}
where $\chi$ is also a function of integration. We can now use this to
eliminate $G$ from the $R_{44}$ component. This leads to the pseudo-wave
equation (\ref{eq:pseudo}):
\begin{eqnarray}
\Box^{'}D&=&D_{11}-2D_{14}+2r^{-1}(D_{1}+D_{4})+r^{-2}(D_{22}+D_{2}\cot\theta)
\nonumber \\
&=&-K+2(F_{14}+2r^{-1}F_{4})+\nonumber \\
& &2r^{-2}\left[\int\{r^{2}(N-2F_{14})+(F_{22}+F_{2}
\cot\theta)\}dr+\chi\right]_{4}.
\end{eqnarray}

To find $G$ we integrate (\ref{eq:geqnapp}) with respect to $\theta$:
\begin{eqnarray}
G&=&r^{-1}\int F_{2}dr+r^{-1}{\rm cosec}\,\theta\int\sin\theta\left[\int r^{2}
(N-2F_{14})dr+r^{2}D_{1}+\chi\right]d\theta\nonumber \\ & &+\nu(r,u)\,
{\rm cosec}\,
\theta
\end{eqnarray}
with $\nu$ being another function of integration. Finally we integrate the
$R_{12}$ component with respect to $r$ and $\theta$ to obtain an expression
for $B$:
\begin{eqnarray}
B&=&{\rm cosec}^{2}\,\theta\int\sin^{2}\theta\left[
-\int\{rL+2r^{-1}(F_{2}-G)\}dr
+F_{2}-G-rG_{1}\right]d\theta\nonumber \\& &+\tau(r,u)\,
{\rm cosec}^{2}\,\theta+
\mu(\theta,u)
\end{eqnarray}
where $\tau, \mu$ are two further functions of integration.

\end{document}